\documentclass[letterpaper]{article} 
\usepackage{aaai24}  
\usepackage{times}  
\usepackage{helvet}  
\usepackage{courier}  
\usepackage[hyphens]{url}  
\usepackage{graphicx} 
\usepackage{natbib}  
\usepackage{caption} 
\usepackage{algorithm}
\usepackage{algorithmic}
\usepackage{amsmath}
\usepackage{amssymb}
\usepackage{multirow}
\usepackage{newfloat}
\usepackage{listings}
\DeclareCaptionStyle{ruled}{labelfont=normalfont,labelsep=colon,strut=off} 
\floatstyle{ruled}
\newfloat{listing}{tb}{lst}{}
\floatname{listing}{Listing}
\nocopyright 

\title{Federated Semi-supervised Learning for Medical Image Segmentation with intra-client and inter-client Consistency}
\author{
    Yubin Zheng\textsuperscript{\rm 1},
    Peng Tang\textsuperscript{\rm 1},
    Tianjie Ju\textsuperscript{\rm 1},
    Weidong Qiu\textsuperscript{\rm 1},
    Bo Yan\textsuperscript{\rm 2}
}
\affiliations{
    \textsuperscript{\rm 1}School of Electronic Information and Electrical Engineering, Shanghai Jiao Tong University\\
    \textsuperscript{\rm 2}Shanghai Chest Hospital\\

    \{zybhk21, tangpeng, jometeorie, qiuwd\}@sjtu.edu.cn, 13817704620@163.com
}

\usepackage{bibentry}

\begin{document}

\maketitle
\begin{abstract}
Medical image segmentation plays a vital role in clinic disease diagnosis and medical image analysis. However, labeling medical images for segmentation task is tough due to the indispensable domain expertise of radiologists. Furthermore, considering the privacy and sensitivity of medical images, it is impractical to build a centralized segmentation dataset from different medical institutions. Federated learning aims to train a shared model of isolated clients without local data exchange which aligns well with the scarcity and privacy characteristics of medical data. To solve the problem of labeling hard, many advanced semi-supervised methods have been proposed in a centralized data setting. As for federated learning, how to conduct semi-supervised learning under this distributed scenario is worth investigating. In this work, we propose a novel federated semi-supervised learning framework for medical image segmentation. The intra-client and inter-client consistency learning are introduced to smooth predictions at the data level and avoid confirmation bias of local models. They are achieved with the assistance of a Variational Autoencoder (VAE) trained collaboratively by clients. The added VAE model plays three roles: 1) extracting latent low-dimensional features of all labeled and unlabeled data; 2) performing a novel type of data augmentation in calculating intra-client consistency loss; 3) utilizing the generative ability of itself to conduct inter-client consistency distillation. The proposed framework is compared with other federated semi-supervised or self-supervised learning methods. The experimental results illustrate that our method outperforms the state-of-the-art method while avoiding a lot of computation and communication overhead.
\end{abstract}

\section{Introduction}
Image segmentation plays an important role in medical image analysis. Accurate and robust segmentation of lesions and organs can assist in diagnosing diseases and formulating treatment plans \cite{masood2015survey}. With the development of computer vision, many deep learning methods are proposed for medical image segmentation and can achieve the state-of-the-art performance with sufficient labeled images \cite{long2015fully,ronneberger2015u,zhao2017pyramid}. However, it is difficult to obtain and maintain a large-scale labeled medical image dataset due to the privacy and sensitive characteristics of medical images. Furthermore, labeling medical images requires a large amount of labour, time and domain expertise \cite{razzak2018deep}. In view of these facts, medical images exist in many isolated medical centers \cite{yang2019federated} and most of the medical images are unlabeled. Federated learning is an effective machine learning approach which allows isolated clients train a global model collaboratively without raw data exchange \cite{mcmahan2017communication}. By applying federated learning to medical image segmentation, each medical center can acquire a machine learning model with enough supervision and keep the privacy of raw data simultaneously \cite{sheller2019multi,rieke2020future,liu2021feddg}.

\begin{figure}[t]
\centering
\includegraphics[scale=0.9]{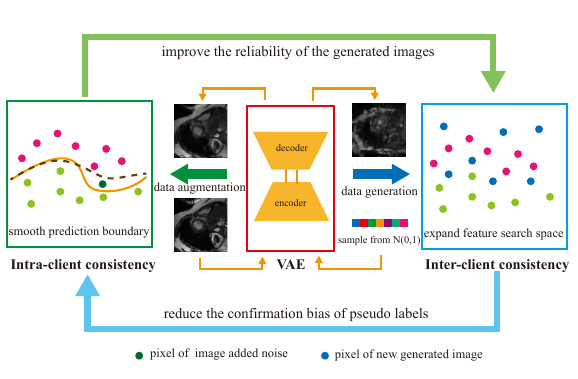}
\caption{The intra-client and inter-client consistency learning mechanism.} \label{fig1}
\end{figure}

Although federated learning is suitable for medical image segmentation, each client possesses a large amount of unlabeled data which can not be utilized in normal supervised federated learning. 
Numerous semi-supervised and self-supervised methods have been introduced to achieve exceptional performance on centralized datasets by leveraging a limited amount of labeled data in conjunction with a large number of unlabeled data \cite{kiyasseh2021segmentation,wang2022rethinking,liu2022acpl,wang2022uncertainty}. However, very few studies considered conducting semi-supervised or self-supervised learning in federated learning scenarios \cite{li2020federated,chang2020synthetic,li2020multi}. By integrating semi-supervised learning (SSL) into federated learning (FL) as federated semi-supervised learning (FSSL), \citeauthor{jeong2021federated} \shortcite{jeong2021federated} firstly presented two application scenarios of FSSL which are Labels-at-Client and Labels-at-Server scenarios. In the Labels-at-Client scenario, both labeled and unlabeled data are available at local clients. In the Labels-at-Server scenario, labeled data are only available at server. Federated contrastive learning (FCL) is a federated self-supervised learning paradigm which applies contrastive learning to federated learning \cite{wen2022federated}. FCL can help clients collaboratively learn the image-level data representations by pre-training.

In this work, we concentrate on the Labels-at-Client scenario of federated semi-supervised learning. We propose a novel federated semi-supervised learning framework for medical image segmentation named FV2IC. A segmentation UNet model \cite{ronneberger2015u} combined with a Variational Autoencoder (VAE) \cite{kingma2013auto} model is collaboratively trained in our framework. Our main idea is illustrated in Figure 1 and mainly includes following four aspects: (1) We consider using VAE model to extract latent global features of labeled and unlabeled medical images. Segmentation models based on CNN backbone such as UNet mostly focus on the features of local pixels in images \cite{li2022semi}. VAE model can extract the global features of medical images which can improve the segmentation performance effectively. (2) We introduce intra-client consistency learning in each client at the data level. The intra-client consistency loss between the images and the augmented images reconstructed by VAE is calculated and added to the supervised segmentation loss to smooth the prediction boundary. (3) The inter-client consistency learning is introduced to urge the predictions of global model to be consistent with the ensemble predictions of clients on the images generated by VAE model which can expand the feature search space. (4) The information of unlabeled images of each client is shared through the VAE model which reduces the interactions between clients and server and avoids a lot of communication overhead.

We apply our framework to two different medical image segmentation tasks and compare our framework with other federated semi-supervised or self-supervised methods. The experiment results indicate that our framework outperforms the state-of-the-art method while reducing the computation and communication cost. In conclusion, we make the following contributions:

\begin{itemize}
    \item We propose a novel federated semi-supervised learning framework for the Labels-at-Client scenario to address the challenges of label scarcity and data island.
    \item An additional VAE model is trained collaboratively to extract global features of the medical images which can enhance the performance of the model effectively.
    \item The intra-client and inter-client consistency learning are proposed with the assistance of VAE model and the effectiveness of them are validated.
    \item Our framework outperforms the state-of-the-art method on two different clinical datasets while utilizing minimal computation and communication resources.
\end{itemize}

\begin{figure*}[h]
\centering
\includegraphics[scale=0.79]{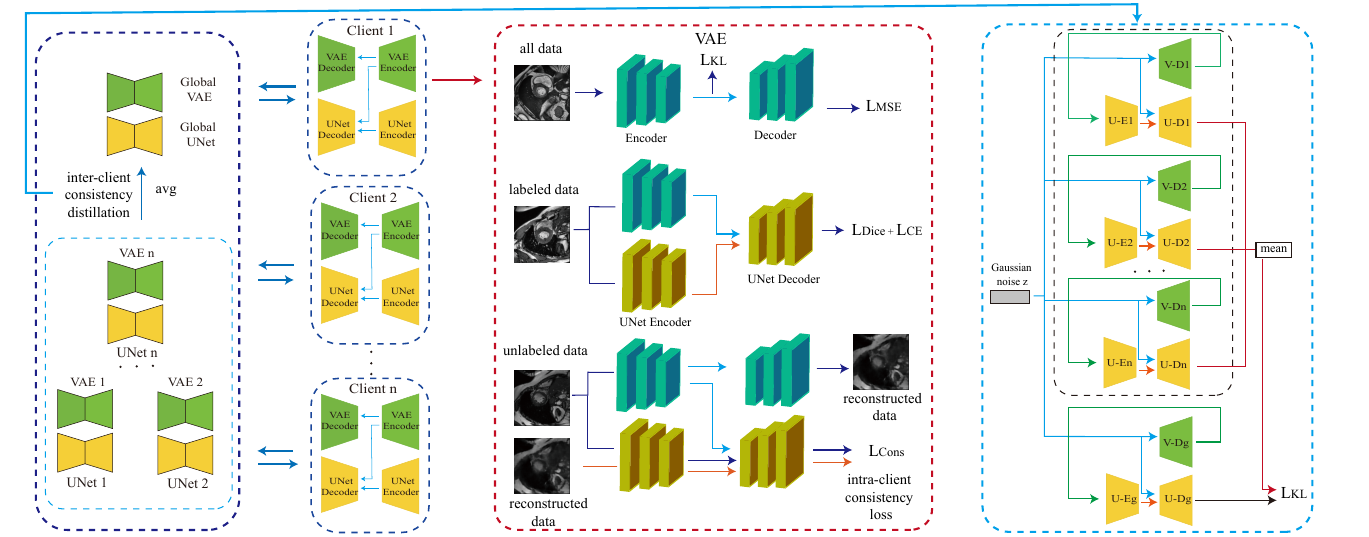}
\caption{The overall architecture of the proposed federated semi-supervised learning framework. The red box shows the training process for each client's local model. The inter-client consistency distillation process is illustrated in the blue box.} \label{fig1}
\end{figure*}

\section{Related Work}
\subsubsection{Semi-Supervised Medical Image Segmentation:} In medical image segmentation, labeling images requires a lot of time, labour and medical expertise. So semi-supervised learning is a favorable paradigm in medical image segmentation. The semi-supervised learning methods can mainly be divided into three categories including pseudo-label learning \cite{bai2017semi,yao2022enhancing}, consistency regularization training \cite{li2018semi,luo2021semi} and entropy minimization \cite{wang2022uncertainty}. Pseudo-label learning aims to employ the trained model to predict pseudo labels of unlabeled data and utilizing them as new labeled samples for further training. For example, \citeauthor{yao2022enhancing} \shortcite{yao2022enhancing} proposed a confidence-aware cross pseudo supervision framework to enhance the quality of pseudo labels derived from unknown distributions. Another approach is to add unsupervised regularization such as consistency loss to the model. For example, \citeauthor{luo2021semi} \shortcite{luo2021semi} creatively introduced a task-level consistency loss and encourage the related data of two tasks in their framework to be consistent. Some research train the model with the principle of low entropy. For example, \citeauthor{wang2022uncertainty} \shortcite{wang2022uncertainty} designed a uncertainty-guided pixel contrastive loss serving as a low entropy regularization which encourages the representations of different geometric transformation to be consistent. Recently, prior knowledge of unlabeled data is used to assist in medical image segmentation. GBDL \cite{wang2022rethinking} utilized a Bayesian deep learning framework which leverages prior knowledge to learn the conditional probability distribution of the model's parameters under the joint distribution of data and labels.

\subsubsection{Federated Semi-Supervised Learning:} A classical federated learning framework comply the agreement that all clients own labels of all training data. For some special scenarios such as medical image analysis, it is unrealistic to require that the data of each client is fully labeled. \citeauthor{jeong2021federated} \shortcite{jeong2021federated} firstly presented two application scenarios in FSSL which are Labels-at-Client and Labels-at-Server scenarios. A FSSL model named FedMatch is also proposed in this work which firstly trains the local model using labeled data and then urges the prediction of the model on unlabeled data to be consistent with other client's ensemble prediction. Several studies have focused on building a federated semi-supervised framework for the pathology classification tasks including FedIRM \cite{liu2021federated} and RSCFed \cite{liang2022rscfed}. FedIRM attempted to align the disease relation between labeled and unlabeled clients while RSCFed employed mean teacher method and update the global model by aggregating multiple sub-consensus models. Meanwhile, some federated semi-supervised learning methods have been proposed for medical image segmentation such as FedCons \cite{yang2021federated} and F2CMT \cite{wen2022federated}.  FedCons is a FSSL framework proposed for COVID region segmentation which leverages the global model to conduct unsupervised consistency learning in unlabeled clients and supervised learning in labeled clients. F2CMT is a FSSL paradigm for medical image segmentation which improves over the mean teacher mechanism with a cross-clients ensemble module. FDPL \cite{qiu2023federated} incorporate a feasible federated pseudo-labeling approach for unlabeled clients which utilize the embedded knowledge learned from labeled clients. One disadvantage of these methods is that the model performance is improved by designing complex federated semi-supervised protocols, but the communication overhead is greatly increased simultaneously. As far as we know, only \cite{yang2021federated,wen2022federated,qiu2023federated} are designed for medical image segmentation tasks. So we also take another federated self-supervised learning paradigm which is federated contrastive learning (FCL) into consideration. \citeauthor{wu2021federated} \shortcite{wu2021federated} proposed a FCL framework for cardiac image segmentation.

\section{Methods}
In this section, we introduce our proposed federated semi-supervised medical image segmentation framework. The framework is illustrated in Figure 2. In the training process of local models of each client, we firstly map the features of images to a low-dimensional space by training a VAE model using all images. To ensure the features obtained by VAE model are relevant to the segmentation task, the VAE encoder is combined with a UNet model to learn the segmentation task using labeled images. All images are input to VAE model to be reconstructed incorporating global data noise. Then the images and the reconstructed images will be input to the segmentation module. The intra-client consistency loss is calculated and added to the supervised segmentation loss. During the aggregation process, each client uploads the parameters of VAE and UNet model to server and sever aggregates the parameters with the FedAvg \cite{mcmahan2017communication} algorithm firstly. And then inter-client consistency distillation is conducted to ensure the predictions of global model to be consistent with the ensemble predictions of clients on the generated images which can alleviate confirmation bias in calculating intra-client consistency loss. 
\subsubsection{Supervised Learning with VAE:} The features of images can be utilized to a low-dimensional space through VAE model. VAE assumes that all samples $\{x_i\}_{i=1}^{N}$ are generated by the discrete latent variable $z$ which subject to the Gaussian distribution. According to the Bayes' rule, we can get $p(x) = \int p(x|z)p(z)dz$. However, the posterior distribution $p(z|x)$ is intractable. VAE adopts an encoder to approximate the posterior distribution which can be denoted by $q(z|x)$. To ensure that the model has generative ability, the distribution $q(z|x)$ is expected to be similar with $p(z)$. So the training goal of VAE is to optimize the output of the encoder $q(z|x)$ to approximate $p(z)$ and maximize the output of the decoder $p(x|z)$ under the distribution $q(z|x)$. The evidence lower bound objective of VAE can be denoted by:
\begin{equation}
    ELBO = \mathbb{E}_{q_\theta(z|x)}log(p_\theta(x|z)) - KL(q_\theta(z|x)||p(z))
\end{equation}
where KL denotes Kullback–Leibler divergence. To maximize $ELBO$, the first item should be maximized and the second item should be minimized. The loss function of VAE can be:
\begin{equation}
\begin{aligned}
    \mathcal{L}_{VAE} &= \mathcal{L}_{KL} + \mathcal{L}_{MSE} \\
                      &= \frac{1}{n}\sum_{i=1}^n(KL(N(\mu_i, \sigma_i^2)||N(0, 1)) + \| x_i - f_v(x_i) \|^2)
\end{aligned}
\end{equation}
where $N(0,1)$ denotes the standard normal distribution, $f_v(x_i)$ denotes the output of VAE model with $x_i$ as input.

By training VAE model, the output of the VAE encoder $z$ can be viewed as the latent low-dimensional features of images. However, the latent features might not be inherently pertinent to the segmentation task. So we combine the VAE encoder with a UNet model to collaboratively learn the segmentation task from labeled images. The training goal is to learn the distribution of labels under the joint distribution of images and their latent features which can be indicated by $p(y_l|x_l,z)$. In this supervised learning process, the UNet encoder receives labeled images $x_l$ to extract features that are further combined with their corresponding $z$ to be input to the UNet decoder. Then the output of the UNet decoder is leveraged to calculate the dice loss $\mathcal{L}_{Dice}$ and cross-entropy loss $\mathcal{L}_{CE}$ with labels $y_l$ to maximize the probability $p(y_l|x_l,z)$. The loss function of supervised training can be:
\begin{equation}
    \mathcal{L}_{sup} = \mathcal{L}_{Dice} + \omega\mathcal{L}_{CE}
\end{equation}
where $\omega$ is a hyper-parameter to balance the dice loss and cross-entropy loss.

\begin{algorithm}[tb]
\caption{Training process of local models}
\label{alg:algorithm}
\textbf{Input}: Clients' datasets $\{D_{lc} \bigcup D_{uc}\}_{c=1}^n$, Clients' VAE encoder $g_e(x)$, Clients' VAE decoder $g_d(z)$, Clients' UNet model $f(x,z)$\\
\textbf{Output}: VAE model parameter $g_c$ and UNet model parameter $f_c$ of each client
\begin{algorithmic}[1] 
\STATE Clients receive $f$ and $g$ from the server.
\FOR{each client $c=1,...,N$ in parallel}
\FOR{iter $j=1,...,iter\_max$}
\STATE Sample batch $b = x_i \in D_{lc} \bigcup D_{uc}$
\STATE $\mathcal{L}_{KL} = \frac{1}{|b|}\sum_{x_i\in b}KL(N(\mu_i, \sigma_i^2)||N(0, 1))$
\STATE $\mathcal{L}_{MSE} = \frac{1}{|b|}\sum_{x_i\in b}\| x_i - g(x_i) \|^2$
\STATE $\mathcal{L}_{VAE} = \mathcal{L}_{KL} + \mathcal{L}_{MSE}$
\STATE Computing gradient of loss function $\mathcal{L}_{VAE}$ and update VAE parameters $g$ by back propagation
\ENDFOR
\FOR{iter $k=1,...,iter\_max$}
\STATE Sample batch $b = x_a \in D_{lc} \bigcup D_{uc}$, sample batch $b_l, y_l = x_i, y_i \in D_{lc}$
\STATE $\mathcal{L}_{Dice} = \frac{1}{|b_l|}\sum_{x_i\in b_l}Dice(f(x_i, g_e(x_i)),y_i)$
\STATE $\mathcal{L}_{CE} = \frac{1}{|b_l|}\sum_{x_i\in b_l}CE(f(x_i, g_e(x_i)),y_i)$
\STATE $\mathcal{L}_{Cons} = \frac{1}{|b|}\sum_{x_a\in b}\| f(x_a,g_e(x_a))-f(g(x_a),g_e(x_a)) \|^2$
\STATE $\mathcal{L}_{Seg} = \mathcal{L}_{Dice} + \omega * \mathcal{L}_{CE} + \lambda(t) * \mathcal{L}_{Cons}$
\STATE Computing gradient of loss function $\mathcal{L}_{Seg}$ and update VAE parameters $g$ and UNet parameters $f$ by back propagation
\ENDFOR
\STATE Upload $g_c$ and $f_c$ to server
\ENDFOR
\end{algorithmic}
\end{algorithm}

\subsubsection{Semi-supervised training with intra-client consistency:} In semi-supervised learning that relies on consistency regularization, consistency losses between original images and augmented images are added to supervised losses to smooth predictions at the data level. A novel intra-client consistency loss is introduced in our FSSL framework. Compared with traditional data augmentation in SSL, we use VAE to reconstruct images and regard the reconstructed images as augmented data. The local VAE model is aggregated through the FedAvg algorithm in each round to update the global VAE model. This aggregation introduces a certain level of global data noise into the images reconstructed by VAE. Consequently, the proposed augmentation strategy approximates the conditional probability distribution $p(y|x_r,z)$ to the distribution $p(y|x_t,z)$ where $x_t$ denotes the original images and $x_r$ denotes the reconstructed images. The intra-client consistency loss can be:
\begin{equation}
\begin{aligned}
    \mathcal{L}_{Cons} &= \frac{1}{n}\sum_{i=1}^n\| f_u(x_i, z_i) - f_u(x_r, z_i) \|^2 \\
                       &= \frac{1}{n}\sum_{i=1}^n\| f_u(x_i, z_i) - f_u(g_d(z_i), z_i) \|^2
\end{aligned}
\end{equation}
where $f_u$ denotes the UNet model with original images $x$ and their latent features $z$ from VAE encoder as input. And $g_d$ represents the decoder of VAE which outputs the reconstructed image $x_r$ by $z$. 

The proposed intra-client consistency can make subsequent inter-client consistency distillation in aggregation process more accurate as the distribution $p(y|x_r,z)$ is trained to be close to the distribution $p(y|x_t,z)$. The loss function of the semi-supervised training can be:
\begin{equation}
    \mathcal{L}_{Seg} = \mathcal{L}_{Sup} + \lambda(t)\mathcal{L}_{Cons}
\end{equation}
where $\mathcal{L}_{Sup}$ is for labeled data, while $\mathcal{L}_{Cons}$ is applied to both labeled and unlabeled data. $\lambda(t)$ is a temperature function of which the value grows gradually with the number of training rounds. The temperature function can balance the supervised and unsupervised loss to prevent gradients from oscillating at the beginning of training. The training process of local models is shown in Algorithm 1.
\subsubsection{Inter-client consistency distillation:} In semi-supervised training, the model will have a certain preference on labeled data which is easy to produce confirmation bias when calculating consistency loss. To overcome this problem, we introduce inter-client consistency distillation in aggregation process. Firstly, each client uploads its local VAE and UNet model to the server. Subsequently, the server aggregates the local models of clients to update the global VAE and UNet model through the FedAvg algorithm. After that, the server randomly samples fixed batches of noises $z$ from Gaussian distribution $N(0,1)$. Then the sampled $z$ is fed to VAE generators and UNet models and two segmentation results $p_c(y|g_d(z),z)$ and $p_s(y|g_d(z),z)$ are calculated. $p_c(y|g_d(z),z)$ is ensembled from the VAE generators and UNet models of each client while $p_s(y|g_d(z),z)$ is from the global VAE generator and global UNet model. Here, $p_c(y|g_d(z),z)$ is close to $p_c(y|x_t,z)$ due to the intra-client consistency mentioned above. Next, the server optimizes the global VAE generator and global UNet model by minimizing the KL divergence between the two distributions.
\begin{equation}
\begin{aligned}
    p_c(y|g_d(z),z) &= \sum_{c=1}^n\frac{|x_c|}{\sum_{c=1}^n|x_c|}f_c(g_{cd}(z),z)
\end{aligned}
\end{equation}
\begin{equation}
p_s(y|g_d(z),z) = f_s(g_{sd}(z),z)
\end{equation}
\begin{equation}
    \mathcal{L}_{KL} = \mathbb{E}_{z\sim\mathcal{Z}}KL(p_c(y|g_d(z),z),p_s(y|g_d(z),z))
\end{equation}
where $|\cdot|$ denotes the size of dataset, f indicates the UNet model and g represents the VAE generator. After inter-client distillation, the server distributes the updated global VAE and UNet model to all clients. By minimizing the KL divergence between two distributions, $p(y|x_t,z)$ incorporates ensemble information which reduces the confirmation bias in calculating intra-client consistency. The aggregation process in server is shown in Algorithm 2.
\begin{algorithm}[tb]
\caption{Aggregation process of the proposed model}
\label{alg:algorithm}
\textbf{Input}: Clients' VAE decoder $g_{cd}(z)$, Clients' UNet model $f(x,z)$\\
\textbf{Output}: Updated VAE model parameter $g$, updated UNet model parameter $f$
\begin{algorithmic}[1] 
\STATE Server aggregate VAE and UNet models through FedAvg to obtain global VAE model and global UNet model $\{g_s,f_s\} = \sum_{c=1}^n\frac{|x_c|}{\sum_{c=1}^n|x_c|}\{g_c, f_c\}$
\FOR{iter $l=1,...,iter\_max$}
\STATE Sample batch $b = z_i \in N(0,1)$
\STATE $y_{ensemble} = \sum_{c=1}^n\frac{|x_c|}{\sum_{c=1}^n|x_c|}\frac{1}{|b|}\sum_{z_i\in b}f_c(g_{cd}(z_i),z_i)$
\STATE $y_s = \frac{1}{|b|}\sum_{z_i\in b}f_s(g_{sd}(z_i),z_i)$
\STATE $\mathcal{L}_{KL} = KL(y_{ensemble}, y_s)$
\STATE Computing gradient of loss function $\mathcal{L}_{KL}$ and update VAE parameters $g$ and UNet parameters $f$ by back propagation
\ENDFOR
\RETURN Updated $g$ and $f$
\end{algorithmic}
\end{algorithm}

\section{Experiments and Results}

\subsubsection{Datasets and Pre-processing:} To evaluate the proposed method, we adopt two segmentation datasets for different tasks which are cardiac images segmentation dataset ACDC 2017 \cite{bernard2018deep} and skin lesions segmentation dataset ISIC 2018 \cite{codella2018skin}.

ACDC 2017 consists cardiac MRI images of 100 patients. The selected volumes are split into training, validation and testing dataset according to the ratio of 70\%, 10\% and 20\%. The 70 cases in training dataset are evenly divided into 10 parts where each part serves as the local data of 10 clients. 20\% of the local data is randomly selected as labeled data while the remaining 80\% is treated as unlabeled data. We adopt min-max normalization to normalize the images and crop the images from middle to a fixed size of $224 \times 224$.

ISIC 2018 includes 2594 dermoscopy images. According to the ratio of 7:1:2, we randomly select 1800 images for training, 260 images for validation and 534 images for testing. As for training dataset, each client is distributed 180 images as local training data and 20\% of them are labeled for semi-supervised experiment setting. We follow \cite{jha2020doubleu} to resize the dermoscopy images to $192 \times 256$.

\subsubsection{Implementation Details and Evaluation Metrics:} We implement our framework using PyTorch library on a NVIDIA RTX3090 GPU. For fair comparisons, all the methods adopt a standard 2D UNet as the backbone. The framework is trained by Adam optimizer with a learning rate of 0.0002 for UNet model and 0.001 for VAE model. Federated training is conducted 200 rounds and each client trains the local model for 1 epoch every round. The batch size is set to 24 including 4 labeled images and 20 unlabeled images. Four metrics are used to evaluate the proposed framework including dice coefficient, jaccard index, sensitivity and accuracy which are introduced in the appendix. As the cardiac images contain three segmentation classes, we average the metrics of three classes to be the final experiment results.

\begin{table*}[t]
\small
\renewcommand\arraystretch{1}
\setlength{\tabcolsep}{7.9pt}
\begin{center}
\begin{tabular}{c|c|c|c|c|c|c|c|c}
\hline
\multirow{2}*{Model} & \multicolumn{2}{c|}{Scan used} & \multicolumn{4}{c|}{Metrics} & \multicolumn{2}{c}{Cost}\\
\cline{2-9}
       & Labeled & Unlabeled & Dice(\%) & Jaccard(\%) & Sensitivity(\%) &  Accuracy(\%) & Time(h) & Comm. (M) \\
\hline
{} & \multicolumn{8}{c}{Dataset 1: ACDC 2017}\\
\hline
FedAvg & $100 \times 10$ & $0 \times 10$ & 89.83 & 83.61 & 91.12 & 99.40 & 1.4 & 416 \\

FedAvg & $20 \times 10$ & $0 \times 10$ & 79.17 & 70.65 & 81.92 & 98.73 & 0.3 & 416 \\
\hline
FedMatch & $20 \times 10$ & $80 \times 10$ & 82.32 & 74.59 & 83.34 & 98.97 & 2.8 & 2496 \\

FCL & $20 \times 10$ & $80 \times 10$ & 82.12 & 74.26 & 83.31 & 98.93 & 3.5 & 1526 \\

FedCons & $20 \times 10$ & $80 \times 10$ & 83.20 & 74.80 & 85.78 & 99.05 & \textbf{2.0} & \textbf{416} \\

FedIRM & $20 \times 10$ & $80 \times 10$ & 84.17 & 76.03 & 85.76 & 99.11 & 2.1 & 534 \\

F2CMT & $20 \times 10$ & $80 \times 10$ & 83.73 & 75.61 & 86.06 & 99.07 & 4.9 & 2288 \\

RSCFed & $20 \times 10$ & $80 \times 10$ & 83.96 & 75.87 & 85.60 & 99.08 & 5.1 & 2376 \\
FDPL & $20 \times 10$ & $80 \times 10$ & 83.50 & 75.29 & 86.26 & 99.07 & 3.2 & 935 \\
\hline
FV2IC(Ours) & $20 \times 10$ & $80 \times 10$ & \textbf{85.57} & \textbf{77.48} & \textbf{86.67} & \textbf{99.18} & 2.6 & 844 \\
\hline
{} & \multicolumn{8}{c}{Dataset 2: ISIC 2018}\\
\hline
FedAvg & $180 \times 10$ & $0 \times 10$ & 88.68 & 81.92 & 89.46 & 95.49 & 2.3 & 416 \\

FedAvg & $36 \times 10$ & $0 \times 10$ & 84.85 & 77.36 & 85.09 & 93.50 & 0.3 & 416 \\
\hline
FedMatch & $36 \times 10$ & $144 \times 10$ & 85.21 & 77.51 & 84.92 & 94.05 & 4.7 & 2496 \\

FCL & $36 \times 10$ & $144 \times 10$ & 85.32 & 77.61 & 84.46 & 94.07 & 6.3 & 1432 \\

FedCons & $36 \times 10$ & $144 \times 10$ & 84.93 & 77.39 & 83.45 & 93.92 & 3.4 & \textbf{416} \\

FedIRM & $36 \times 10$ & $144 \times 10$ & 85.65 & 78.05 & 86.04 & 94.31 & \textbf{3.3} & 527 \\

F2CMT & $36 \times 10$ & $144 \times 10$ & 85.46 & 77.93 & 85.53 & 94.15 & 8.8 & 2288 \\

RSCFed & $36 \times 10$ & $144 \times 10$ & 85.53 & 77.94 & 85.51 & 94.23 & 9.2 & 2376 \\
FDPL & $36 \times 10$ & $144 \times 10$ & 85.81 & 78.31 & 85.99 & 94.38 & 5.8 & 948 \\
\hline
FV2IC(Ours) & $36 \times 10$ & $144 \times 10$ & \textbf{86.56} & \textbf{78.94} & \textbf{87.03} & \textbf{94.40} & 5.2 & 836 \\
\hline
\end{tabular}
\label{tab1}
\caption{Quantitative comparison between our method and other federated semi-supervised methods on ACDC 2017 and ISIC 2018 dataset. Last column is the communication cost of the methods for each round.} \label{tab1}
\end{center}
\end{table*}


\subsubsection{Compared with Other FSSL Methods:} 
 We compared our framework with seven advanced federated semi-supervised or self-supervised methods including FCL, FedMatch, FedCons, FedIRM, F2CMT, RSCFed and FDPL. 
 
 Table 1 shows the quantitative comparison of our framework and other methods on ACDC 2017 and ISIC 2018. All the methods obtain the performance improvement through semi-supervised or self-supervised learning manners compared with the FedAvg method using only 20\% labeled images in each client. The FedIRM, F2CMT, RSCFed and FDPL achieve higher performance than FedMatch, FCL and FedCons. It substantiates that more complicated protocols of federated learning or more advanced semi-supervised learning manners such as mean teacher are helpful to improve the performance of the methods. Our framework outperforms all the methods in four metrics. Compared to the best-performing model, our model still achieved the improvement of 1.40\% in cardiac segmentation task and 0.35\% in skin lesions segmentation task on the dice coefficient. 
 
 It is worth noting that some methods which attain better results such as F2CMT and RSCFed have high training and communication cost. In our framework, the semi-supervised learning including intra-client and inter-client consistency learning is conducted with the assistance of a VAE model which can extract the information of unlabeled images in each client. Only a lightweight VAE model requires to be transmitted between clients and server which avoids taking up a large number of computation and communication resources. Furthermore, cardiac segmentation is a multi-class segmentation task, whereas skin lesion segmentation is a single-class segmentation task. Our model achieves the best performance for both single-class and multi-class segmentation tasks. This further proves the robustness of our model in representing medical images.

\subsubsection{Effectiveness under Different SSL Settings:} We perform a study on effectiveness of our approach under different labeled data percentage. Our framework is compared with the FedIRM which achieves the best performance in six compared methods and fully-supervised UNet using only 20\% annotated images as trained data. The experimental results are shown in Figure 3. The top red dotted line in the figure indicates the upper bound trained with 100\% labeled data in each client. It can be observed that the performance gap between our method and fully-supervised UNet increases with less labeled data. This indicates that our method utilizing labeled data in an efficient manner especially under less labeled data circumstances. Furthermore, the performance of our model is consistently higher than the FedIRM method. An important observation is that the dice coefficient of our method closely approaches the upper bound when the labeled images account for 60\%. It illustrates that our federated semi-supervised learning manner is effective and robust.

\begin{figure}[htp]
\centering
\includegraphics[scale=0.48]{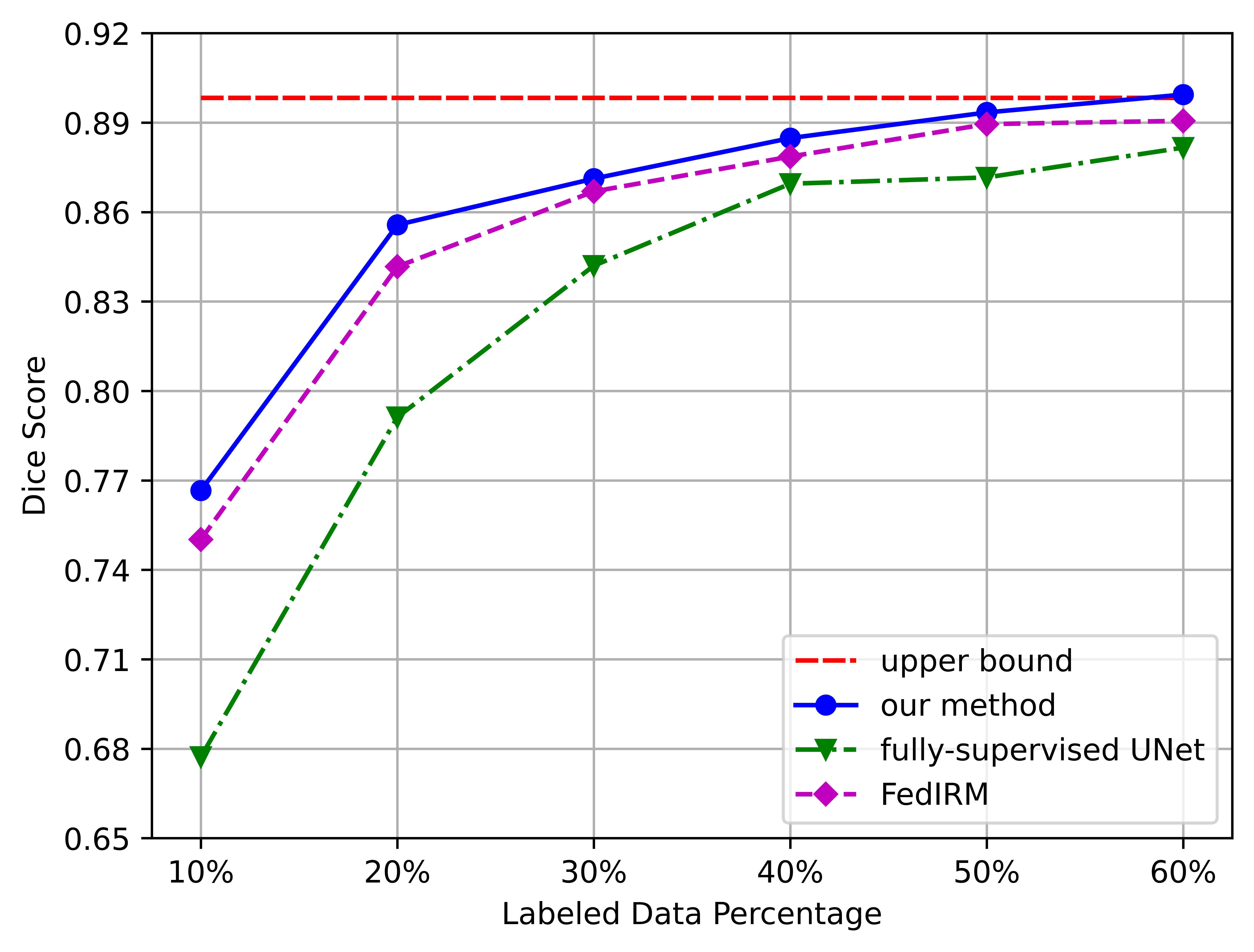}
\caption{The cardiac segmentation performance of our approach with different ratio of labeled data.} \label{fig3}
\end{figure}

\subsubsection{Visual Comparisons:} We visualize the segmentation results of different methods on ACDC and ISIC dataset in Figure 4. Some methods with better performance are chosen including F2CMT, FedIRM, RSCFed and FDPL.
The two cases above are the results of the segmentation on ACDC dataset while the two cases below are the results on ISIC dataset. The green, orange and blue regions denote the right ventricle, myocardium and left ventricle in cardiac MRI images. In the first case, there are all some false positives (red box) appeared in the segmentation results of four compared methods. However, our method has no false positives and possesses higher overlap ratio. In the remaining cases, the segmentation results own many false negatives (red box) compared with the GroundTruth. As for the result of our method, the false negatives are fewer and more details are preserved. These experimental results illustrate the effectiveness, robustness and generalization of our proposed method.


\begin{figure}[htp]
\centering
\includegraphics[scale=0.8]{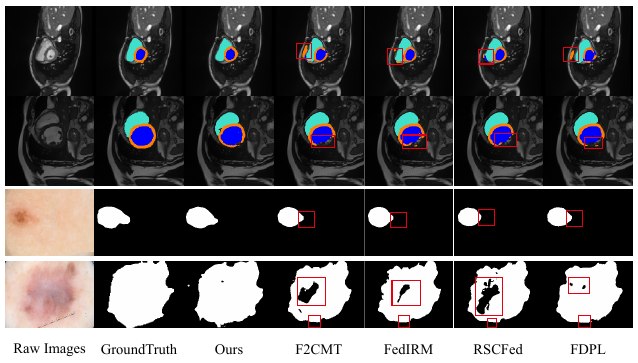}
\caption{Visual comparison of the segmentation results predicted by different methods on ACDC and ISIC datasets.} \label{fig3}
\end{figure}

\begin{table*}[htp]
\renewcommand\arraystretch{1}
\setlength{\tabcolsep}{5pt}
\begin{center}
\begin{tabular}{c c c c c c|c c c c}
\hline
\multirow{2}*{$\mathcal{L}_{dice}$} & \multirow{2}*{$\mathcal{L}_{CE}$} & Gaussian & VAE & VAE & Inter-client & \multirow{2}*{Dice(\%)} & \multirow{2}*{Jaccard(\%)} & \multirow{2}*{Sensitivity(\%)} & \multirow{2}*{Accuracy(\%)}\\
& & aug. & aug. & feature & distillation & & & & \\
\hline
\checkmark & & & & & & 79.00 & 70.37 & 82.38 & 98.70\\
\checkmark & \checkmark & & & & & 79.17 & 70.65 & 81.92 & 98.73\\
\hline
\checkmark & \checkmark & \checkmark & & & & 80.79 & 72.24 & 84.97 & 98.79 \\
\checkmark & \checkmark & & \checkmark & & & 83.01 & 74.79 & 85.61 & 99.03 \\
\checkmark & \checkmark & \checkmark & & \checkmark & & 81.88 & 73.60 & 84.60 & 98.91\\
\checkmark & \checkmark &  & \checkmark & \checkmark & & 84.68 & 76.37 & 86.65 & 99.11\\
\hline
\checkmark & \checkmark & & \checkmark & \checkmark & \checkmark & \textbf{85.57} & \textbf{77.48} & \textbf{86.67} & \textbf{99.18}\\
\hline
\end{tabular}
\label{tab3}
\caption{Ablation study of the proposed federated semi-supervised learning framework on ACDC dataset.}
\end{center}
\end{table*}

\subsubsection{Ablation Study:} Table 2 presents the results of ablation study of our framework on ACDC dataset. The addition of the cross-entropy loss $\mathcal{L}_{CE}$ results in a 0.17\% improvement in the dice coefficient. In order to validate the effectiveness of our implemented data augmentation method involving VAE-based image reconstruction, we conduct a comparison with a conventional data augmentation technique that involves adding Gaussian noise to images. Our data augmentation method conducted by VAE achieves better performance in all four metrics and especially has 2.3\% improvement in the dice coefficient.

We propose the use of features extracted by VAE encoder to assist UNet model in segmentation task. To verify the effectiveness of this component, we separately add VAE features to the models using two different data augmentation methods. By adding VAE features, two models using different data augmentation methods all obtain improvement in four metrics. The application of VAE features combined with data augmentation through the VAE model yields a notable 1.67\% enhancement in the dice coefficient.

The inter-client consistency distillation module is integrating to the model incorporating VAE augmentation and VAE features to validate its effectiveness. After integrating inter-client consistency distillation, the dice coefficient of the model has increased by 0.89\% and other three metrics has also increased which demonstrates that the inter-client consistency distillation can enhance the effectiveness of our semi-supervised learning approach.

\begin{figure}[h]
\centering
\includegraphics[scale=0.6]{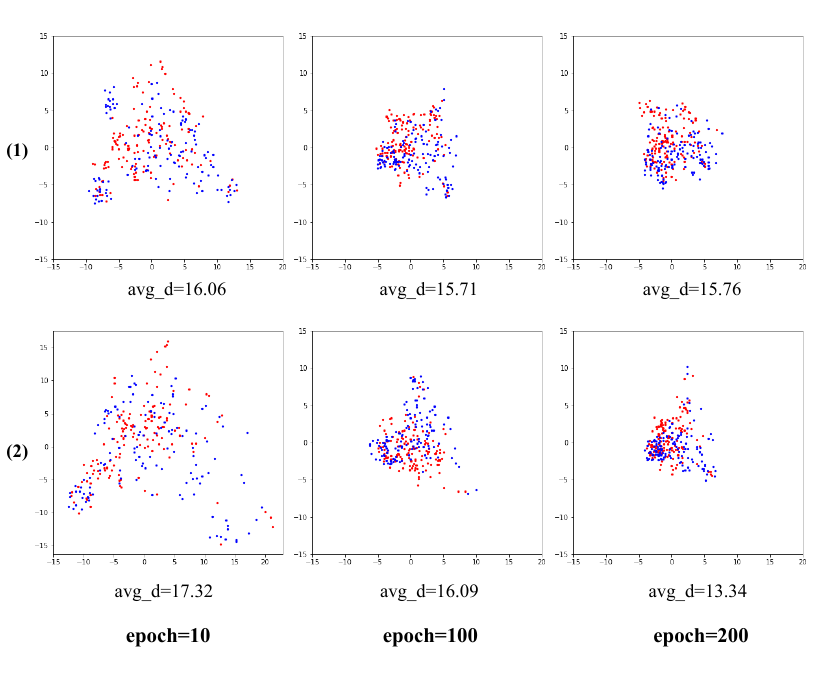}
\caption{The comparison of features $z$ between the combined VAE-UNet model (second row) and a single VAE model (first row). Red points denote the samples with dice coefficient higher than 0.9 while blue points with dice coefficient lower than 0.9. avg\_d is the average Euclidean distance of $z$ vectors.} \label{fig4}
\end{figure}

\subsubsection{Adaptive Adjustment of VAE Features for Segmentation Task:} We consider the encoder of VAE as a feature extractor and combine it with a UNet model to learn the segmentation task collaboratively. The purpose is to make the features extracted by VAE adapt to the segmentation task. So the training goal of the model is changed to maximize the probability of $p(y|x,z)$ where $z$ is the feature extracted by VAE. Labels $y$ can be regard as highly abstracted images of the original images. The features of a few dimensions in vector z have a crucial impact on the segmentation task. This will consequently improve the similarity of features $z$. We visualize the features $z$ output by VAE encoder on ACDC test dataset. Figure 5 shows the experimental results. It is evident that points, particularly the red points, become more distinctly clustered as the number of epochs increases in the combined VAE-UNet method (second row). The average Euclidean distance of $z$ vectors in combined VAE-UNet is also smaller than that of a single VAE model at the 200th epoch. These results demonstrate the adaptive adjustment of VAE features for the segmentation task in our method.

\subsubsection{Convergence Analysis:} To explore the convergence of the model, we visualized supervised and unsupervised losses on ACDC dataset. We randomly selected two clients including client 1 and client 6. It can be seen from Figure 6 that both the supervised dice loss and unsupervised intra-client and inter-client consistency loss exhibit satisfactory convergence as the number of training rounds increases. The decrease of the inter-client consistency loss demonstrates that the global model can be optimized in uncertain feature space. This will greatly improve the generalization ability of the model and reduce the induction bias.

\begin{figure}[t]
\begin{minipage}[b]{.49\linewidth}
  \centering
  \centerline{\includegraphics[width=4.8cm]{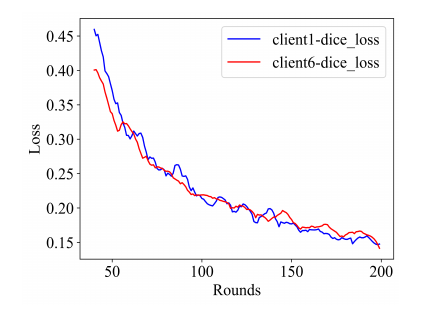}}
  \centerline{(a) Supervised loss}\medskip
\end{minipage}
\begin{minipage}[b]{.49\linewidth}
  \centering
  \centerline{\includegraphics[width=4.8cm]{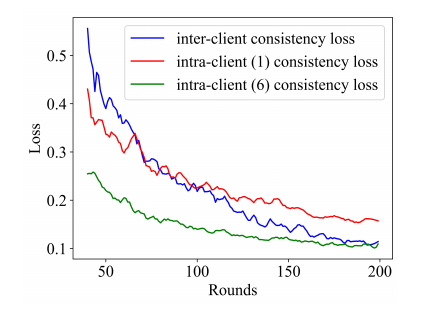}}
  \centerline{(b) Unsupervised loss}\medskip
\end{minipage}
\caption{Convergence of supervised and unsupervised loss}
\label{fig5}
\end{figure}

\section{Conclusion}
In this paper, we present a novel federated semi-supervised learning framework for medical image segmentation. The intra-client and inter-client consistency learning are proposed to smooth predictions at the data level and alleviate the confirmation bias of local models. They are conducted with the support of a generative VAE model. VAE model can also extract the favourable features of medical images which can assist in segmentation task. The experimental results demonstrate that our approach outperforms the state-of-the-art method and each component of our framework is effective. In the future, we will focus more on the safety of our framework. Some methods such as differential privacy and homomorphic encryption will be added to our model to prevent the malicious agents from stealing private information of images.

\bibliography{aaai24}

\end{document}